\documentclass[sn-nature,iicol]{sn-jnl}% Default with double column layout

\usepackage{graphicx}%
\usepackage{multirow}%
\usepackage{amsmath,amssymb,amsfonts}%
\usepackage{amsthm}%
\usepackage{mathrsfs}%
\usepackage[title]{appendix}%
\usepackage{xcolor}%
\usepackage{textcomp}%
\usepackage{manyfoot}%
\usepackage{booktabs}%
\usepackage{algorithm}%
\usepackage{algorithmicx}%
\usepackage{algpseudocode}%
\usepackage{listings}%
\usepackage{soul} % strikeout
\newcommand{\K}{$\mathrm{K^{+}}\;$}
\newcommand{\Ca}{Ca$^{2+}$}

\begin{document}

\title
[Lipid-mediated hydrophobic gating in the BK potassium channel]
{Lipid-mediated hydrophobic gating in the BK potassium channel}

\author[1]{\fnm{Lucia} \sur{Coronel}} % \email{iiauthor@gmail.com}
\author[2]{\fnm{Giovanni} \sur{Di Muccio}} \email{giovanni.dimuccio@uniroma1.it}
\author[3]{\fnm{Brad} \sur{Rothberg}} 
\author[2]{\fnm{Alberto} \sur{Giacomello}}%\email{iiiauthor@gmail.com}
\author[1]{\fnm{Vincenzo} \sur{Carnevale}}\email{vincenzo.carnevale@temple.edu}

\affil[1]{\orgdiv{Institute for Computational Molecular Science and Institute for Genomics and Evolutionary Medicine and Department of Biology}, \orgname{Temple University}, \orgaddress{\street{1925 N. 12$^{\mathrm{th}}$ St.}, \city{Philadelphia}, \postcode{19122}, \state{PA}, \country{USA}}}

\affil[2]{\orgdiv{Department of Mechanical and Aerospace Engineering}, \orgname{Sapienza University of Rome}, \orgaddress{\street{Via Eudossiana, 18}, \city{Rome}, \postcode{00184}, \country{Italy}}}

\affil[3]{\orgdiv{Department of Medical Genetics and Molecular Biochemistry}, \orgname{Temple University}, \orgaddress{\street{3500 N. Broad St.}, \city{Philadelphia}, \postcode{19140}, \state{PA}, \country{USA}}}


\abstract{
The large-conductance, calcium-activated potassium (BK) channel lacks the typical intracellular bundle-crossing gate present in most ion channels of the 6TM family. 
This observation, initially inferred from Ca$^{2+}$-free-pore accessibility experiments and recently corroborated by a CryoEM structure of the non-conductive state, raises a puzzling question: how can gating occur in absence of steric hindrance? To answer this question, we carried out molecular simulations and accurate free energy calculations to obtain a microscopic picture of the sequence of events that, starting from a Ca$^{2+}$-free state leads to ion conduction upon Ca$^{2+}$ binding.
Our results highlight an unexpected role for annular lipids, which turn out to be an integral part of the gating machinery. Due to the presence of fenestrations, the ``closed'' Ca$^{2+}$-free pore can be occupied by the methyl groups from the lipid alkyl chains. This dynamic occupancy triggers and stabilizes the nucleation of a vapor bubble into the inner pore cavity, thus hindering ion conduction.
By contrast, Ca$^{2+}$ binding results into a displacement of these lipids outside the inner cavity, lowering the hydrophobicity of this region and thus allowing for pore hydration and conduction. 
This lipid-mediated hydrophobic gating rationalizes several seemingly problematic experimental observations, including the state-dependent pore accessibility of blockers.
}
 

\maketitle
    
\newpage

\section{Introduction} 
High-conductance $\mathrm{K^+}$ channels (BK channels) are finely tuned molecular machines that gate intracellular $\mathrm{K^+}$ flux in response to $\mathrm{Ca^{2+}}$ binding and depolarization. 
They are ubiquitously expressed in nerves and muscles and fulfill diverse biological functions that range from hearing and neurosecretion to muscles contractions. 
Each subunit of the tetrameric BK channel consists of a voltage sensing domain (VSD), a pore-gate domain (PGD), and a cytosolic tail domain (CTD) (Fig.~\ref{fig:1}).
The PGD enables permeation of selected ionic species across the cell membrane, while the VSD and CTD allosterically modulate pore gating. 
The transmembrane region of each monomer comprises seven helices, S0 through S6, with S1-S4 responsible for detecting membrane depolarization (VSD) and S5-S6 constituting the PGD. 
The loop between S5 and S6 forms the selectivity filter (SF), which accommodates at least two \K ions simultaneously~\cite{doyle1998} (Fig.~\ref{fig:1}b).
The PGD water-filled central cavity, is also the binding site for some cationic pore blockers~\cite{Zhou2001}. The CTD contains binding sites for intracellular \Ca and triggers channel activation; 
BK channels can be activated independently via \Ca binding to the CTD or in response to membrane depolarization~\citep{yang2015bk,miranda2013state}.

An interesting property of BK PGD, compared to other potassium channels, is a pronounced kink of the pore-lining S6 helix between Ala316 and Ser317~\cite{Zhou2015}. 
A rotation at this kink is part of the conformational transition that underlies in channel gating~\citep{Chen2011,Chen2014,wu2009}.
Despite the fact that several experimental structures in both conductive and non-conductive states have clarified the roles played by various components, several crucial aspects of the gating mechanism remain unclear. 
In particular, although the \Ca-free conformation (PDB 5TJI~\citep{Tao2017}) presents a slightly narrower pore radius compared to the Ca$^{2+}$-bound one (PDB 5TJ6~\cite{Tao2017}), neither structure shows a pore that is sufficiently occluded by the pore-lining helices to hinder diffusion of waters and ions to and from the central cavity. 
In the \Ca-free conformation, the radius of the pore is still large enough (as large as $\sim 7 \,$\AA) to allow occupancy by multiple water layers, ions and other molecules~\cite{Wilkens2006,Jia2018,Tao2017,Hite2017}.

One of the molecular mechanisms that has been proposed to explain the transition toward a non-conductive state is hydrophobic gating~\citep{ARYAL2015,Jia2018}.
This mechanism relies on dewetting phenomena that can occur within narrow hydrophobic pores or channels, owing to a combination of spatial restriction and unfavorable interaction between water molecules and exposed non-polar residues~\citep{beckstein2004not,giacomello2020bubble, guardiani2021unveiling, paulo2023hydrophobically}.
Since the BK inner PGD core presents a hydrophobic region just below the SF, in particular near residue Leu-312, the hypothesis that hydrophobic gating is at work in BK channels is supported by several computational and experimental studies~\cite{ARYAL2015,Hite2017,Jia2018}.
This mechanism seems to consistently explain how the pore may remain accessible to relatively large hydrophobic molecules while relatively small ions become impermeant. However, the BK PGD volume seems to exceed the critical radius, length and hydrophobicity demonstrated to be needed to allow and maintain a bubble nucleation required to disrupt ion permeation~\cite{ARYAL2015, rao2019heuristic, trick2014designing, paulo2023hydrophobically, giacomello2020bubble, huang2023hydrophobic}.
Hence, as illustrated in this work, additional ingredients are required to mediate the pore complete dewetting for such pore.

We explore the molecular basis of pore-gating in BK channel via a quantitative investigation of conductive and non-conductive states through extensive Molecular Dynamics (MD) simulations that combine equilibrium and enhanced sampling calculations.
Our analysis unveils additional physical and 
thermodynamic features underlying the gating process,
as compared to previous computational studies on ion channels~\citep{Jia2018,lynch2020water,gu2023central}.
In particular, our results strongly emphasize the fundamental, yet unforeseen, role of membrane lipids in the dewetting transition. 
Lipids turn out to be crucial players in BK inactivation dynamics, since their binding and intrusion through lateral fenestrations -- only present into the \Ca-free conformation -- increases the pore inner cavity hydrophobic character, markedly triggering dewetting.
Finally, these annular lipids play also a crucial role in coupling PGD and CTD: \Ca binding to the CTD causes a displacement of these lipid molecules that makes them unable to bind to the fenestration and thus inhibit ion conduction. Overall, our results reveal an unsuspected regulatory mechanisms governing the BK channel activation relying on the interplay between \Ca binding, PGD-lipid interactions, and hydrophobic gating.

\section{Methods}
\begin{figure}[t]\centering
    \includegraphics[width=0.47\textwidth]{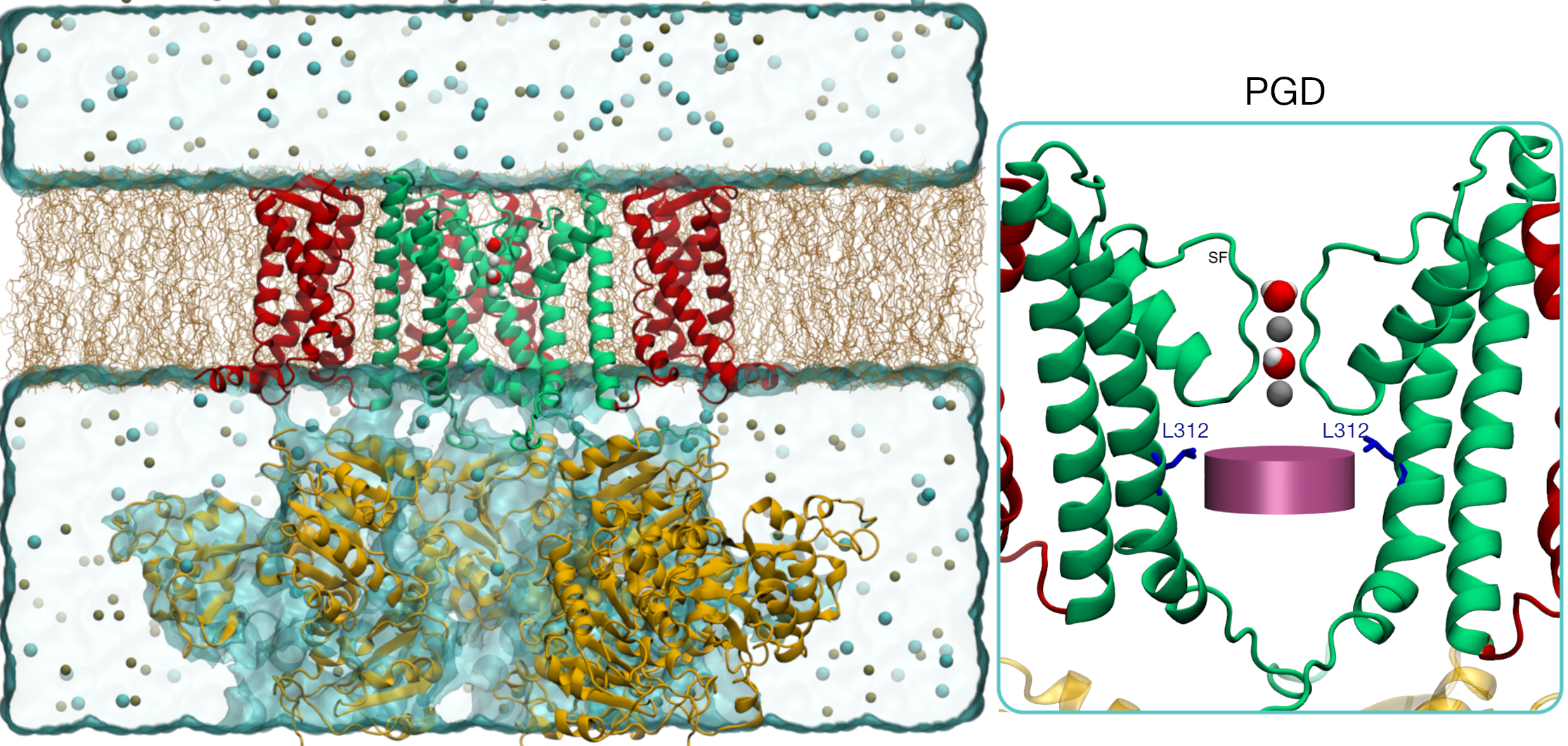}
    \caption{
    Side view of BK channel. 
    Left: BK channel is as cartoon: the S1-S4 helices (VSD) are highlighted in red, the S5–S6 helices in green (PGD), and the intracellular portion (CTD) in yellow. Membrane lipids (ochre) and ions are shown as thick lines and spheres, respectively. The water molecules of the selectivity filter are shown in space filling representation. For clarity, only three subunits are shown. 
    Right: BK channel pore gate domain (only two subunits shown for clarity), highlighting the K$^+$ ions (gray sphere) and waters (S1(top)–S4(bottom)) occupying the  SF and the deep pore volume (pink cylinder). The side chain of residue L312 is shown in blue. 
    \label{fig:1}
    }
\end{figure}

\textbf{MD setup and equilibrium simulations of the full length and truncated BK channel.} The BK channel is studied in its \Ca-bound and \Ca-free state (ID PDB 6V38 and ID PDB 6V3G respectively ~\citep{Tao2019}), via MD simulations of the full length as well as of the transmembrane domains, separately.
In the latter case the protein is truncated at the ASN328 residues. 
The molecular systems are assembled using the CHARMM-GUI web service~\cite{jo2008}, specifically the channel proteins are embedded in a lipid bilayer of 1-palmitoyl-2-oleoyl-sn-glycero-3-phosphocholine (POPC) with a number of ions adjusted to obtain electrical neutrality at salt concentration of 0.15M. The temperature and pressure are maintained via Nosé–Hoover temperature coupling method, (tau-t=1 ps), and semi-isotropic Parrinello–Rahman method (tau-p=5 ps), respectively~\cite{Parrinello1981,Nose1983,Nose1984,Hoover1985}. Electrostatic interactions are treated by the PME algorithm~\cite{1995PME}. Two potassium ions and two water molecules are allocated in the selectivity filter (TVGYG), in position S1-S3 and S2-S4 respectively. This arrangement is kept constant by applying harmonic restraints to the ions and waters positions. 
The equilibrium all-atoms simulations are carried out using HPC facility, patching GROMACS 2021.4 and PLUMED 2.8, 
and using CHARMM36 forcefield~\cite{brooks2009charmm} with TIP3P water~\cite{jorgensen1983} under periodic boundaries conditions. 
After the initial minimisation and equilibration of $\sim 2\,ns$, following the standard CHARMM-GUI protocol, the systems is evolved using a time step of $2\,$fs and collecting trajectories for a total of $1000\,$ns. 
We utilized the HOLE software~\citep{HOLE1996} to calculate the protein pore lumen dimensions. For each conformation, we computed the average over time and determined the error through block analysis.
We designate a specific region of interest within the deep pore, 
that we refer to as deep pore volume (DPV), as follows: 
a cylindrical shape with a radius of $0.7\,$nm, and the cylinder's central axis passes through the center of mass (COM) of the SF. 
This cylinder begins at a distance of $1.2\,$nm away from the COM and extends for a length of $0.5\,$nm, ending in proximity of the kink of S6 (Fig ~\ref{fig:1}. Finally, to estimate the number and type of molecules contained within this region we utilized VMD~\cite{HUMP96} scripts.
The \K ions permeation free energy profile, sampled along the trajectory according to the canonical weights prescribed by the Boltzmann distribution, can be expressed as:
$-k_B T \log\left( \eta(z)\right)$.
Here, $\eta(z)$ represents the histogram of the \K ions vertical coordinate along the channel axis during the simulation, within $0.7\,$nm from the axis. The constants used are $k_B=1.987\cdot10^{-3}$ kcal/(mol~$\cdot$ K) and $T=303.1 \,$ K.
To estimate the errors we performed block analysis.

\bigbreak
\noindent
\textbf{Enhanced RMD Simulations and pore filling free energy profiles.}
From the $\mathrm{\mu s}$ equilibrium MD of the truncated pore, we extracted a series of configuration to better sample the wet/dry transition of the pore, in the presence or absence of anular lipids. 
For each extracted configuration, we used restrained molecular dynamics simulations (RMD)~\cite{maragliano2006temperature} to extract the (water) filling free energy profile of the pore, using NAMD~\cite{phillips2005scalable} and the Colvars module~\cite{fiorin2013using}.
For each configuration, the protein backbone and the phosphorus of POPC molecule heads are constrained.
Simulations are performed using different water models:
SPC/E~\cite{takemura2012water} and TIP3P~\cite{jorgensen1983}.
The two models reproduce differently the surface tension as compared to the experimental one~\cite{vega2007surface}, a fundamental parameter in bubble nucleation and hydrohpobic gating~\cite{giacomello2020bubble}.
Hence, to test the robustness of our result, part of the simulations are replicated using both the models, see Fig.~\ref{fig:tip3pvsspce}.
SPC/E water is used in combination with Amber force field
(ff15ipq force ﬁeld~\cite{debiec2016further} for the protein and the Lipid17 force-ﬁeld~\cite{dickson2014lipid14} for the phospholipids), 
while TIP3P is used in combination with CHARMM36~\cite{huang2013charmm36}.
RMD are conducted restraining the number of water molecules inside a confined region of the space, by adding an harmonic potential to the Hamiltonian of the system
\begin{equation}\label{eq:restraint}
H_N(\boldsymbol{r},\boldsymbol{p})=H_0(\boldsymbol{r},\boldsymbol{p})+\frac{k}{2}\left(N-\tilde{N}(\boldsymbol{r})\right)^2\,,
\end{equation}
where $\boldsymbol{r}$ and $\boldsymbol{p}$ are the positions and momenta of all the atoms, respectively, $H_0$ is the unrestrained Hamiltonian, $k$ is a harmonic constant which was set to $1\,$kcal/mol, 
$N$ is the desired number of water molecules in a control box, and 
$\tilde{N}$ is the related counter for the actual number of water molecules in the system at every step. A Fermi distribution with Fermi parameter equal to $3\,$\AA~ is used to smooth the borders of the box and make the collective variable continuous.
The protocol is implemented in NAMD by using the Volumetric map-based variables of the Colvars Module~\cite{fiorin2013using}.
The centre and size of the counting box was set equal to the center and the {\it minmax} size of the rings composed by the residues Thr287 and Glu321 computed by VMD~\cite{HUMP96}.
The water molecules affected by the counting box are represented in VDW spheres in Fig.~\ref{fig:4}, within the dashed magenta box.
Each filling state, corresponding to a single point of the profiles reported in Fig.~\ref{fig:4}a and Fig.~\ref{fig:fe_nolipid} was simulated for $4\,$ns. Production averages are performed after $2\,ns$, discarding the transient; representative traces reporting the number of water molecules inside the control box along the simulation, are reported in Fig.~\ref{fig:rmd_traj}. 
Each trajectory is saved every $20\,$ps, while the number of water molecules inside the box is saved every $1\,$ps. 
The free energy gradient (the mean force) for each point is computed as 
$-k\langle \tilde{N} - N \rangle$ ~\cite{maragliano2006temperature}; standard error is computed via block average, with a block length of $100\,ps$. Then, the free energy profiles for are obtained by thermodynamic integration of multiple RMD simulations (40 points). Integration is performed using the Euler method; the standard errors are propagated correspondingly.

\section{Results and Discussion}

\begin{figure*}[ht!]
\centering
    \includegraphics[width=0.95\textwidth]{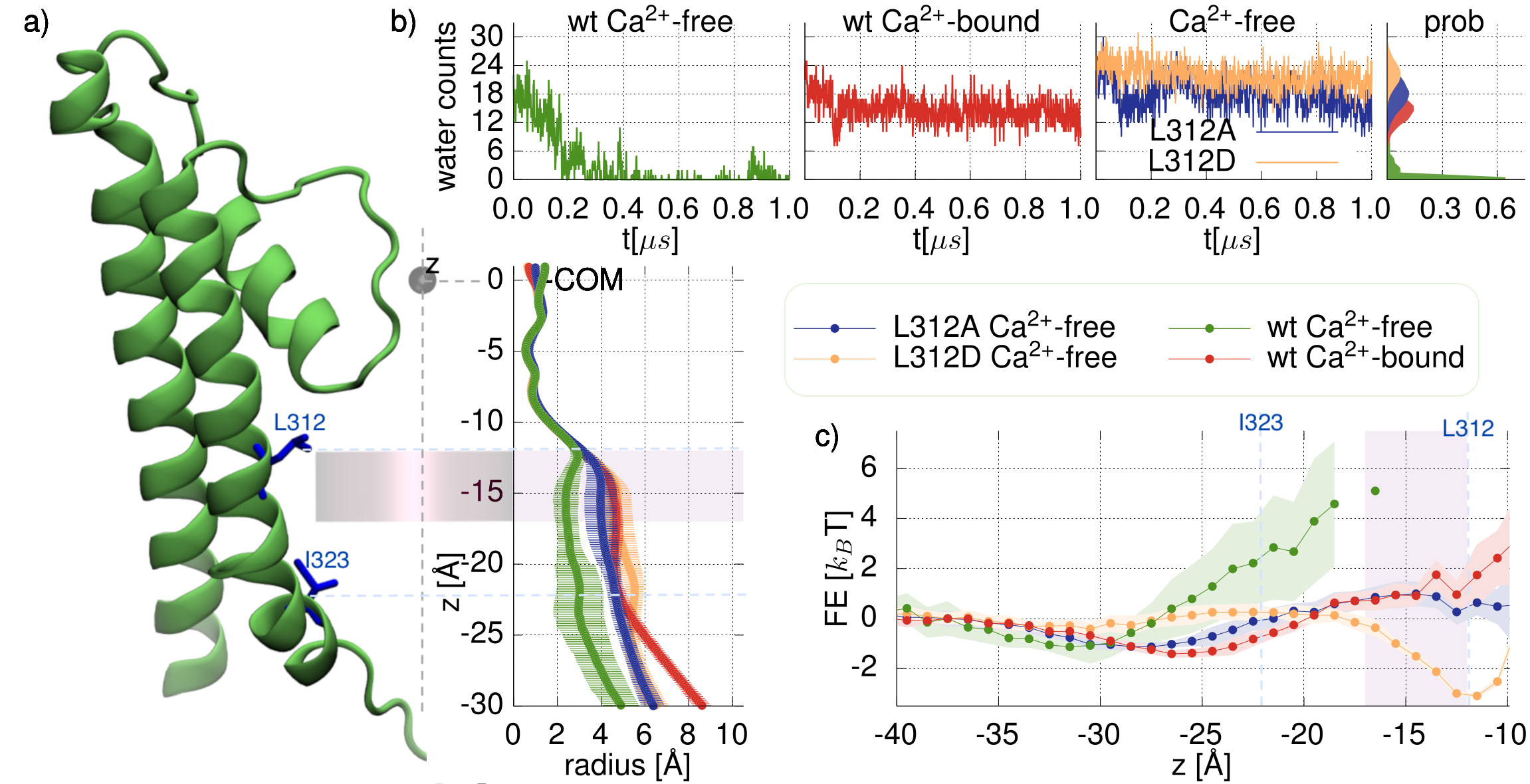}
    \caption{ 
    a) Average radius profile for wild-type $\mathrm{Ca^{2+}}$-bound (red), wild-type $\mathrm{Ca^{2+}}$-free (green), mutant L312A $\mathrm{Ca^{2+}}$-free (blue) and L312D $\mathrm{Ca^{2+}}$-free (orange). $z=0$ corresponds to the SF center of mass (COM) and the pink shaded area indicates the location of the DPV. 
    b) Counts of water molecules within the DPV. Shown in vertical to the right is the distribution probability computed for $t \ge 200 \,$ ns.
    c) Ion permeation free energy profiles; the zero is set at the free energy value corresponding to $z=-40\,$\AA. The DPV is highlighted be the pink shaded area.
    \label{fig:2}
    }
\end{figure*}

\textbf{Pore hydration controls ion occupancy.} 
To gain insight toward the molecular factors affecting ion permeation,  we collected an extensive set of $1\mu s$-long MD trajectories of the human BK channel in both the \Ca-bound and \Ca-free states, along with single residue mutants investigated in  previous electrophysiological studies~\citep{Chen2014,Carrasquel2014}, L312A and L312D.

The conductive (\Ca-bound) and non-conductive (\Ca-free) states show markedly different radius profiles in the region delimited on one side by the cytosolic mouth and on the other by the SF (Fig.~\ref{fig:2}a).
The radii of the \Ca-bound and \Ca-free states begin to branch out $\sim 12\,$\AA~ below the center of mass of the selectivity filter, and keep diverging up to the S6 kink, up to $z=-17 \,$\AA, where the difference between the two pore radii achieves a maximum (Fig.~\ref{fig:dpv}), where the radii of the \Ca-bound and \Ca-free state are, respectively, $r_o=4.7 \,$ \AA~ and $r_c=2.4 \,$ \AA. 
We refer to this region, highlighted in pink in Fig.~\ref{fig:2}, 
as the deep pore volume (DPV). 
%Into the DPV, the \Ca-bound radius is always larger than the \Ca-free. 
Although small, the pore radius of the \Ca-free configuration is twice the radius of a water molecule, {\sl i.e.}, the section is large enough to accommodate two to four water molecules.
However, 
looking at the ions permeation free energy profiles along the pore axis (Fig.~\ref{fig:2}b), 
the \Ca-free state show a marked difference throughout the transmembrane region with respect to the \Ca-bound state, larger than the one expected for a similar channel radius reduction and in presence of water. 
%Indeed, the barrier ratio between the two state is $\sim 4$ at the DPV entrance $z=-15\,$\AA; assuming that the pore conductance is proportional to the rate constant $\kappa \propto e^{-\beta \Delta F^\dagger}$, with $\Delta F^\dagger$ being the free energy barrier to enter the DPV, it would results a drop in conductance for the \Ca-free state greater than $98\%$ as compared to the \Ca-bound state. 
Note that in the \Ca-free configuration the DPV ion occupancy is not observed with statistically significant data, due to dewetting phenomena, which suggests that larger barriers may actually occur.

The presence of significant ionic permeation free energy barrier in the \Ca-free conformation despite the absence of a large steric hindrance was previously observed in other works~\cite{Jia2018,gu2023central} and was shown to be related to hydrophobic gating through the central cavity of the pore, as indeed observed also in our simulations.
To quantify the DPV hydration level, we measured the number of water molecules occupying this volume as a function of time.
Consistently with the previously reported observation~\cite{Jia2018}, the pore of the non-conductive state reaches a state devoid of water molecules after a certain time (Fig.~\ref{fig:2}c): 
the \Ca-free state exhibits large fluctuations in the number of water molecules, decreasing to zero after $200\,$ns.
Differently, for the \Ca-bound state the hydration level remains relatively constant along the trajectory.
The difference between the two conformations is dramatically summarized by the distributions of water counts computed after $200\,$ns:  the averages of the water counts are $0.8$ and $14.3$ for the \Ca-free and \Ca-bound states, respectively (Fig.~\ref{fig:2}c).

%To test the hypothesis that the dewetting of PGD is a fundamental feature affecting the pore ion permeability, 
Then, we investigated the behavior of two L312 mutants, L312A and L312D.
Alanine substitution (L312A) partially conserves the hydrophobic character of the pore while slightly increasing the pore radius, while aspartate substitution (L312D) represents an isosteric mutation, albeit it introduces a negative charge which renders the site highly hydrophilic. Experimentally, the L312A mutant shows a median voltage of activation close to $0\,$mV even at 0M~\Ca, while L312D mutation results into a constitutively open channel~\citep{Chen2014}. 

The four symmetry-related residues at position 312 are located on S6 at the DPV top edge (Fig.~\ref{fig:1} and Fig.~\ref{fig:2}a).
Remarkably, our simulations show that both mutations makes the L312(A,D) \Ca-free conformation similar to the WT \Ca-bound one, displaying similar values of the pore radius in the DPV, absence of ion permeation free energy barriers, and similar hydration levels (Fig.~\ref{fig:2}).
Specifically, the L312A pore radius in the DPV (blue profile) closely resembles that of the \Ca-bound state (red), while it converges toward the \Ca-free (green) value of $5-6\,$\AA~ at the cytosolic mouth. Similarly, the L312D pore radius (orange) is increased compared to the wild type \Ca-free state and it is larger than the \Ca-bound state one. 
The increased DPV radius of the mutants, compared to the WT \Ca-free conformation, leads to the water molecules count increase, with an observed average count of $18 \pm 3 $ and $22 \pm 3$ for L312A and L312D respectively, {\sl i.e.}, similar to the \Ca-bound value (Fig.~\ref{fig:2}c). The ultimate effect of the DPV radius enlargement is a dramatic decrease of the ions permeation free energy barrier: the free energy profile of the (\Ca-free) L312A is statistically indistinguishable from that of the \Ca-bound conformation (Fig.~\ref{fig:2}b). These results are consistent with the experimental observation that the G-V curve of L321A in absence of \Ca~ is significantly left-shifted whereas L312D lead to constitutively open channel. Furthermore, they corroborate the previously posited hypothesis of hydrophobic gating, {\sl i.e.}, that ion permeability depends strongly on the hydration state of a specific region of the pore, namely the DPV. 

However, the results so far obtained from both wild type and mutation simulations are insufficient to fully elucidate the mechanism underlying the DPV dewetting. Indeed, by computing the wetting and drying free energy profile of the inner pore cavity of the initial \Ca-free configuration, see Fig.~\ref{fig:fe_nolipid}, we did not observe any stable minimum in the dry state, as it would be expected for an hydrophobically gated pore.
This is not particularly surprising, since -- although small -- the DPV radius, length and hydropathy of the \Ca-free configuration seems to be larger than the critical size and length usually reported to be needed to trigger a stable hydrophobic gating~\cite{beckstein2004not,ARYAL2015,rao2019heuristic,trick2014designing,paulo2023hydrophobically}. 
As described in the next paragraph, a better inspection of the dewetted region seems to solve this gap, bringing to light the indispensable role of an unsuspected player: the anular lipids.

%%%%%%%%%%%%%%%%%%%%%%%%%%%%%%%%%%%%%%%%%%%%%%%%%%%%%%%%%%%%%%%%%%%%%%

\bigbreak
\noindent
\textbf{Lipid tails induce dewetting.} 
A close inspection of our molecular simulation trajectories reveals that DPV dewetting occurs exclusively when the pore is partially occupied by the hydrophobic lipid tail methyl groups. These intrusions are made possible by the presence of four symmetry-related protein cavities -- hereafter referred to as fenestrations -- connecting the pore lumen to the bulk section of the lipid bilayer.
These fenestrations are present only in the \Ca-free structure as a result of a kinked conformation of S6 (Fig.~\ref{fig:helices} and Fig.~\ref{fig:helices_details}).
Furthermore, the CryoEM structures (PDB ID 6V38) 
display electron density in the fenestrations attributed to annular
lipids~\cite{Hite2017,Tao2017,Tao2019,Fan2020}.

\begin{figure}[t]
\centering
 \includegraphics[width=0.47\textwidth]{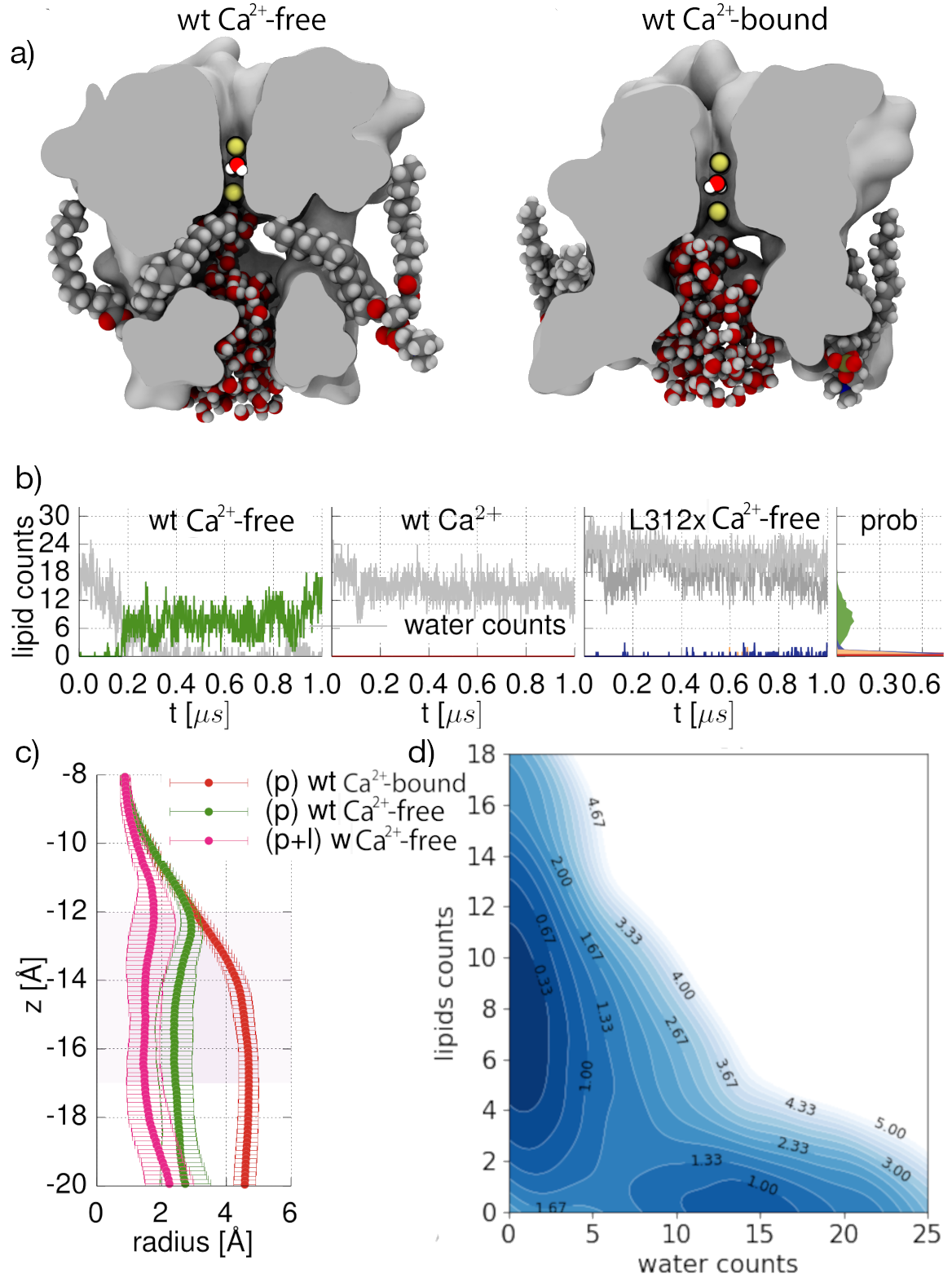}
\caption{
    a) Snapshots from the \Ca-free and \Ca-bound trajectories. 
    To highlight the shape of fenestrations, the location of the intruding lipids and the pore occupancy by water molecules, we show a section of the PGD molecular surface with the section-cut plane oriented along the C$_4$ axis of symmetry. 
    b) Lipid carbon atoms counts within the DPV, for wild-type \Ca-free (green), wild-type \Ca-bound (red), mutant L312A \Ca-free (blue) and L312D (orange) state. For comparison, water molecules counts are shown in gray. c) Average radius profiles: the pink line shows the profile for the wild-type \Ca-free conformation when lipid atoms are considered in the pore radius calculation (the DPV is the highlighted by the pink background). 
    d) 2D histograms of the water and lipids atom counts in the wild-type \Ca-free state. For visual clarity, a negative logarithmic scale is used for plotting the values obtained through kernel density estimation and the global minimum is set to zero.
 \label{fig:3}
}
\end{figure}

We analyzed the position of lipids carbon atoms along the trajectory,
in particular, focusing on those occupying the DPV 
(Fig.~\ref{fig:3}a). 
Despite the fact that the initial configuration of both \Ca-free and \Ca-bound systems have no lipids tails inside the fenestrations, after $200\,ns$ the DPV of the \Ca-free configuration starts to be occupied by up to $8 \pm 3$ carbon atoms.
As the lipid carbon atoms fill the DPV in the \Ca-free configuration, 
the water molecules count decreases steadily until nearly hits zero due to the decreased accessible volume in the pore lumen;
the DPV radius ($ z \sim -15 \,$ \AA ) decreases by $40 \% $, passing from $r_{p} =2.4 \,$ \AA~ to $ r_{p+l}=1.5 \,$ \AA~ (Fig.~\ref{fig:3}b-pink line). 
By contrast, no lipids are observed in the pore of \Ca-bound state nor in the L312A mutant of the \Ca-free configuration, showing hydrated pores throughout the trajectory. 
Overall, water and lipid pore occupancies are mutually exclusive, as suggested by the anticorrelation shown in Fig.~\ref{fig:3}c, 
which highlights two distinct DPV states: a wet one without lipids, and a dry one with lipids.

Since the transition between the wet and dry state of a nanocavity is a rare event~\cite{giacomello2012cassie}, 
the quantitative estimation of the relative probabilities of these states would require sampling over longer time-scales.
Hence, we employed a different strategy, 
exploiting enhanced Restrained MD simulations~\cite{paulo2024voltage, paulo2023hydrophobically}, 
to establish how the relative stability of the two states -- and their separating free energy barrier -- changes with the number of lipids carbon atoms inside the central cavity, keeping the protein structure constrained.
Indeed, despite previous studies have shown a correlation between the dehydration level of the pore and protein structural features, 
{\sl e.g.}, the orientation of Phe residues in the central cavity or the radius shrinking~\cite{gu2023central}, the complete dewetting of the region just below the SF was rarely reported~\cite{Jia2018}.

\begin{figure*}[t]\centering
 \includegraphics[width=0.9\textwidth]{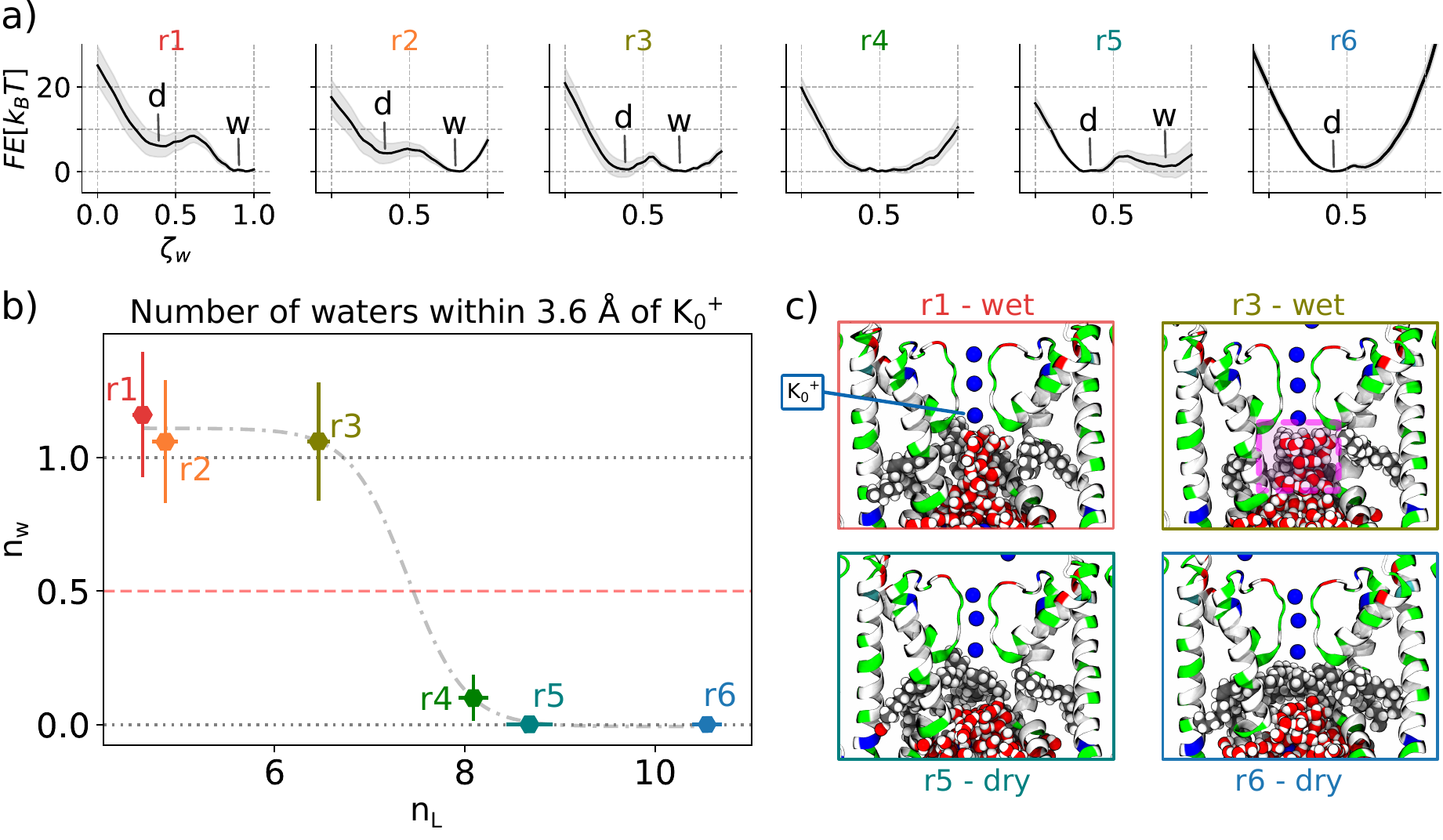} 
 \caption{
 \textbf{a)} Water filling free energy profiles for different pore replicas, 
 having different configurations and average numbers of lipids inside the lumen.
 The free energy profiles are obtained by thermodynamic integration of multiple Restrained Molecular Dynamics (RMD) simulations 
 with SPC/E water model, and using as a collective variable the number of water molecules inside a control box surrounding the hydrophobic region of the pore lumen, indicated in magenta in panel c, top right.
 The profiles of the different replicas are ordered by decreasing number of lipid carbons inside the control box, {\sl i.e.}, 
 replica $r1$ has the largest number of lipids and $r6$ the lowest.
 \textbf{b)} Number of water molecules within $3.6$~\AA~of the innermost potassium in the SF, here indicated as $K_0^+$, for each global minimum of the profiles of panel a. This monitors if -- for the equilibrium state -- water molecules get in touch with the SF, a fundamental step for the ions conduction though the pore.
 \textbf{c)} Last frame of the RMD simulations corresponding to the global minima of four representative replicas.
 The pictures clearly show that water molecules get in touch with the SF in $r1$ and $r3$ but not in $r5$ and $r6$. 
 }
 \label{fig:4}
\end{figure*}

Via RMD enhanced simulations, we computed the free energy as a function of the number of waters molecules in the cavity for the \Ca-free channel (Fig.~\ref{fig:4}). 
As expected, the water filling free energy profiles display two minima, corresponding to the wet and dry (meta)stable states. Calculations performed on multiple molecular configurations, each characterized by a different number of lipid carbon atoms inside the DPV, show how the relative stability of the two states depends on the number of lipids through the cavity: 
below a certain number of lipid carbon atoms, the stable state is the dry one (replica r1, r2); around a value $n_L \sim 7$, the two states becomes equiproblable (r3, r4);
after this value the dry state becomes the most probable (r5, r6).
It should be noted that, in the dry state, water molecules are not anymore in contact with the potassium $K^+$ ions in the SF, substantially preventing their solvation and, consequently, the conductance of the ions through the pore. 
This effect is completely missing in the \Ca-free structure when no lipid are intruded inside the DPV (Fig.~\ref{fig:fe_nolipid}), 
with such ``closed'' structure presenting only one free energy minima in the wet state. 
To compute the free energy profiles shown in Fig.~\ref{fig:4} the SPC/E water model is used, as we found that the TIP3P water model is not able to well reproduce the metastabilities arising in hydrophobic gating (Fig.~\ref{fig:tip3pvsspce}) probably due to the lower surface tension of the TIP3P model~\cite{vega2007surface} as compared to the experimental one. Anyway, regardless of the water model, the main finding -- that the lipids induces the central pore cavity dewetting -- is confirmed.

Overall, these results support the conclusion that the 
\Ca-free configuration -- {\sl i.e.} when no lipids occupy the DPV -- is constitutively hydrated, thereby allowing ion permeation. Thus, binding and intrusion of lipids tails through the lateral fenestrations is seemingly a necessary step for transitioning toward a truly non-conductive configuration. In other words, the reported hydrophobic gating of the BK channel~\cite{Jia2018} is mediated by the entrance of lipid tails through the fenestrations.
In this regards, the behaviour of the L312A mutant is of particular interest: despite the fact that the hydrophobic character of the DPV is preserved by the mutation, the pore remains wet and permeable to ions throughout the simulations, since lipids do not show significant affinity for the fenestrations in that case and do not occupy the DPV.

%%%%%%%%%%%%%%%%%%%%%%%%%%%%%%%%%%%%%%%%%%%%%%%%%%%%%%%%%%%%%%%%%%%%%%
%%%%%%%%%%%%%%%%%%%%%%%%%%%%%%%%%%%%%%%%%%%%%%%%%%%%%%%%%%%%%%%%%%%%%%
%%%%%%%%%%%%%%%%%%%%%%%%%%%%%%%%%%%%%%%%%%%%%%%%%%%%%%%%%%%%%%%%%%%%%%
%%%%%%%%%%%%%%%%%%%%%%%%%%%%%%%%%%%%%%%%%%%%%%%%%%%%%%%%%%%%%%%%%%%%%%

\begin{figure}[t]\centering
 \includegraphics[width=0.47\textwidth]{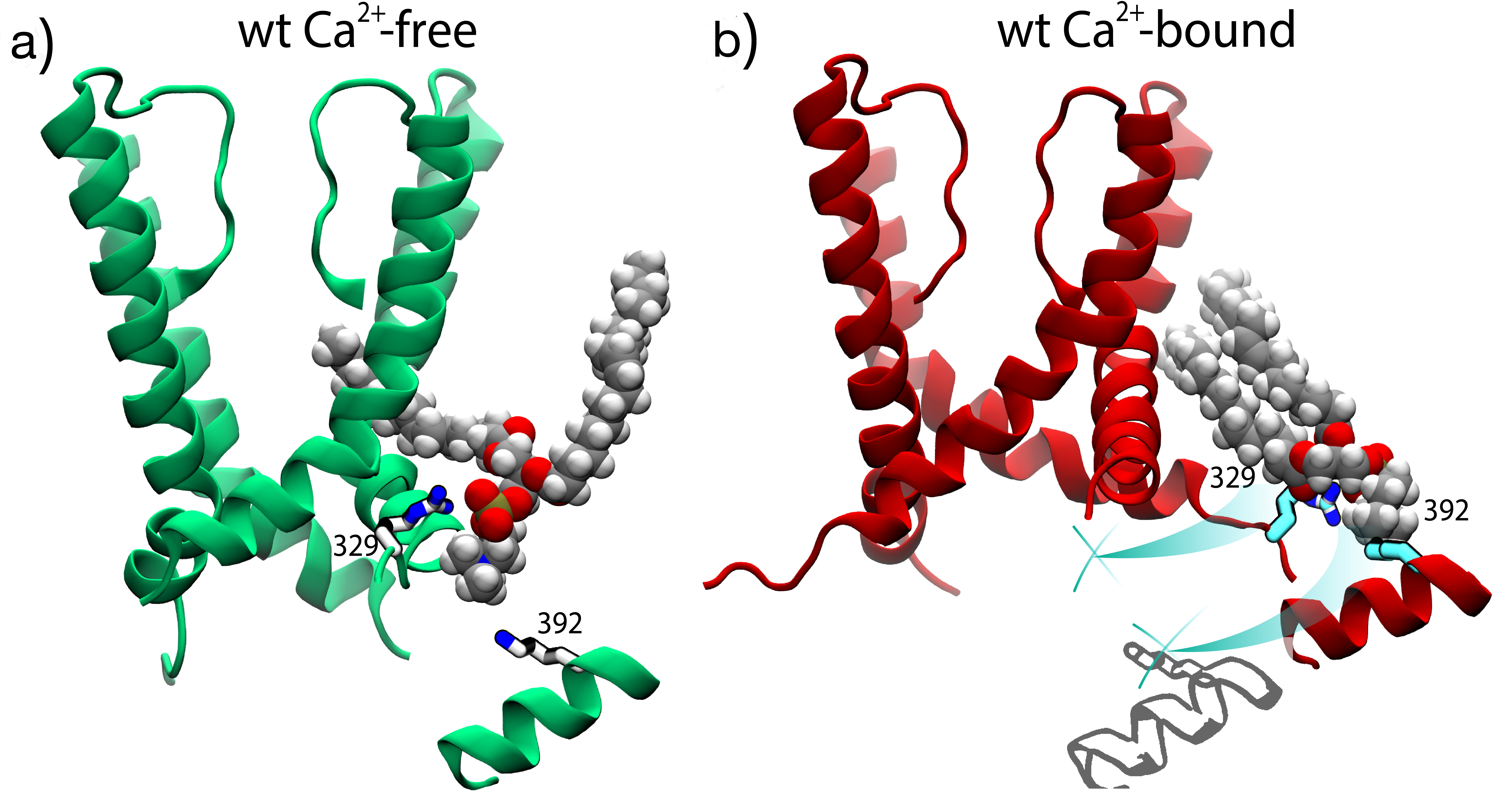}   
 \caption{Trajectory snapshots in the Ca$^{2+}$-free (green) and Ca$^{2+}$-bound (red). Lys392 and Arg329 are shown as sticks. The turquoise lines indicate the displacements of those residue between the two states. For clarity, the Ca$^{2+}$-free helix is shown in gray in b). Shown are the lipids which are, simultaneously, within 5 \AA~ of S6, Lys392, and Arg329. }
 \label{fig:fig5}
\end{figure}

\bigbreak
\noindent
\textbf{Lipids may be involved in allosteric coupling. } 
In addition to producing a significant left-shift of the G-V curve, almost all single residue mutations at position 312 abolish \Ca~ sensitivity ({\sl i.e.}, the G-V curve at 0~\Ca concentration is the same as that obtained at 85 $\mu$M \Ca, see ref.~\cite{Chen2014}).
This suggests the possibility that the lipid molecules involved in pore dewetting contribute also to the allosteric coupling between the CTD and PGD~\cite{Horrigan2002,yang2015bk,Chen2014,tang2014,Hite2017,ceballos2019,rockman2020}.
To test this hypothesis, we investigated how the annular lipids (Fig.~\ref{fig:3}d) interact with the channel 
(Fig.~\ref{fig:l-p-interaction}).
Our results revealed a pattern of interactions between the lipid head groups and residues belonging to the CTD and the S6 C-term region.
Specifically, the interactions with Arg-329 and Lys-392 emerge as key to anchoring the lipid head groups in their respective positions (Fig.~\ref{fig:fig5}). 
Remarkably, this is consistent with previous experiments showing that mutation of the $^{329}$RKK$^{331}$ ring to hydrophobic amino acids promotes the open conformation, shifting the voltage dependence of activation to the negative direction by up to 400 mV~\citep{Tian2019}.
While both Arg-329 and Lys-392 interact with the lipid head groups in both the \Ca-free and \Ca-bound states, only in the former the tail groups are close enough to the fenestration to penetrate into the pore lumen. In particular, Lys392 from the $\alpha$B helix undergroes a significant displacement ($\sim$9$\,$\AA,  Fig.~\ref{fig:displ}) upon \Ca binding. 
As a result, the lipid molecule anchored to this residue is rigidly moved away from the PGD (Fig.~\ref{fig:fig5}). 
Furthermore, the CTD conformational change causes an increase of the S6 bundle which results in the closing of the fenestrations (Fig.~\ref{fig:helices}). 
Overall, the conformational changes caused by \Ca binding affect the positioning of annular lipids, lending confidence to the notion of their involvement in allosteric coupling.

\section{Conclusions}

In this work we investigated the BK channel gating mechanism using equilibrium and enhanced sampling molecular dynamics simulations. 
Our findings align with the view that BK channel gating is controlled by the hydration level of the central cavity and show that, in particular, the area beneath the 
selectivity filter plays a crucial role in gating as it is differentially hydrated in the calcium-free and calcium-bound states. Notably, it is within this 
zone that dewetting phenomena occur due to the intrusion of lipid tail methyl groups, which stably occupy the fenestration throughout our simulations. 

The occurrence of lipid interactions exclusively in the calcium-free state highlights their critical role in facilitating the observed dewetting transition, which in turn inhibits ion permeability. Furthermore, our findings suggest that the stability of the dry state depends on the level of intrusion of the lipid tails into the cavity. 
We determined that under around a certain threshold of carbon atoms inside the DPV ($n_L$=7), both the dry and wet states are equally probable, but a higher number of lipids tips the balance in favor of the dry state, and conversely, a lower number favors the wet state. 
Interactions between lipids and specific amino acids within the S6 helix C-terminal and the C-terminal domain (CTD) appear vital for favoring the penetration of lipids and for keeping the lipid heads close to the central cavity. This supports the notion of a dynamic, functional connection between lipid-CTD interactions and channel activity.

Our study examines the effects of mutations in key areas, offering valuable insights into the complex 
dynamics that regulate the hydrophobic dewetting transition in BK channels. 
Our results are in accordance with 
comprehensive mutagenesis studies~\citep{Chen2014,Carrasquel2014}, 
where a notable correlation was observed between the side chain hydropathy 
-- {\sl i.e.,} the hydration propensity~\cite{di2021characterizing}- of the innermost PGD region and the channel open probability~\citep{Chen2014,Carrasquel2014}. 
Specifically, hydrophilic mutations (S, T, N, D, Q) of L312 not only induce a significant left shift of the activation curve, indicating a stabilization of conductive state, but also significantly reduce \Ca dependence. Specifically, in the calcium-free state, mutations L312A and L312D, which are experimentally known to stabilize the conductive state, closely mimic the geometrical properties and hydration levels of the calcium-bound state, effectively blocking lipid entry. These findings underscore the importance of L312 in adjusting the channel's ion permeability and contribute to a deeper understanding of the mechanisms controlling BK channel states.

In conclusion, our comprehensive analysis of BK channel gating mechanisms through equilibrium and enhanced sampling molecular dynamics simulations provides significant insights into the interplay between lipid interactions, hydration levels, and channel activity. 
By elucidating the critical role of the selectivity filter area, lipid occupancy, and specific amino acid interactions in modulating the gating process, our work sheds light on the intricate molecular dynamics that 
underpin ion permeability and channel functionality. 
The exploration of mutations further enriches our 
understanding of the structural determinants critical for channel behavior, highlighting the nuanced balance between the dry and 
wet states that dictate the gating mechanism. Our findings not only advance the current understanding of BK channel gating suggesting an active role of lipids, but also pave the way for future studies aimed at exploring therapeutic interventions targeting these mechanisms.

\section{Acknowledgement}
This research includes calculations carried out on HPC resources supported in part by the National Science Foundation through major research instrumentation grant number 1625061 and by the US Army Research Laboratory under contract number W911NF-16-2-0189 and by EuroHPC PRACE awarded project ElectroHG (EHPC-REG-2022R03-112) on the LUMI@CSC infrastructure. V.C. acknowledges support by the National Institute for General Medical Science through grant 3R01GM093290.
This project has received funding from the European Research Council (ERC) under the European Union’s Horizon 2020 research and innovation programme (grant agreement No 803213).

\bibliography{bibliography}% common bib file

%%%%%%%%%%%%%%%%%%%%%%%
% SM
\newgeometry{
 a4paper,
 total={180mm,267mm},
 left=25mm,
 right=25mm,
 top=35mm,
 bottom=30mm,
 }
\newpage
\bigskip
\setcounter{page}{1}
\renewcommand{\thefigure}{S\arabic{figure}}
\renewcommand{\thefigure}{S\arabic{figure}}
\setcounter{figure}{0}

\center
{\Large
SUPPLEMENTARY MATERIALS for \\ \vspace{0.2cm}
Lipid-induced hydrophobic gating in BK channels
}

%\bigskip
%\bigskip

%\noindent{\bf Contents:} \linebreak

%\begin{tabular}{ll}
%Material and Methods.                     & p. 2  \\
%\\
%Supplementary Note S1: DPV
%& p. 2  \\
%Supplementary Note S2: Bah
%& p. 3  \\

%Supplementary Figure S-RMDTRAJ: Pore filling during the RMD trajectory
%& p. xx  \\
%\end{tabular}

\newpage
%\newpage

%%%%%%%%%%%%%%%%%%%%%%%%%%%%%%%%%%%%%%%%%%
% SUPP NOTES
%%%%%%%%%%%%%%%%%%
%\section*{Supplementary Note}
%\subsection*{S1.   }

%\subsection*{S2.  }

%%%%%%%%%%%%%%%%%%%%%%%%%%%%%%%%%%%%%%%%%%
% SUPP FIGURES
%%%%%%%%%%%%%%%%%%
\newpage
\section*{Supplementary Figures}

\subsection*{FIGURE S1. - DPV  }
\begin{figure}[h]
\centering
a)\includegraphics[width=0.45\textwidth]{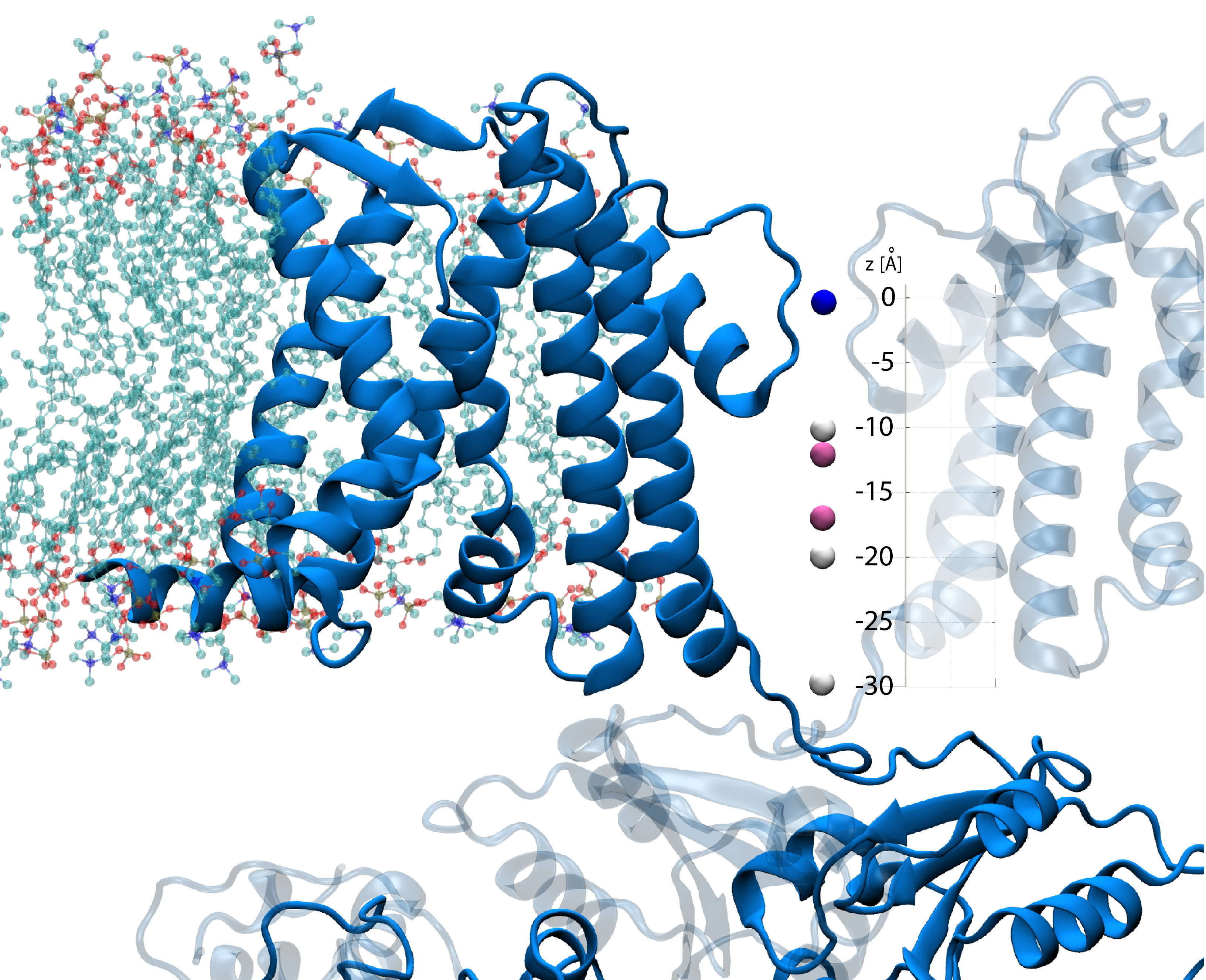}
b)\includegraphics[width=0.45\textwidth]{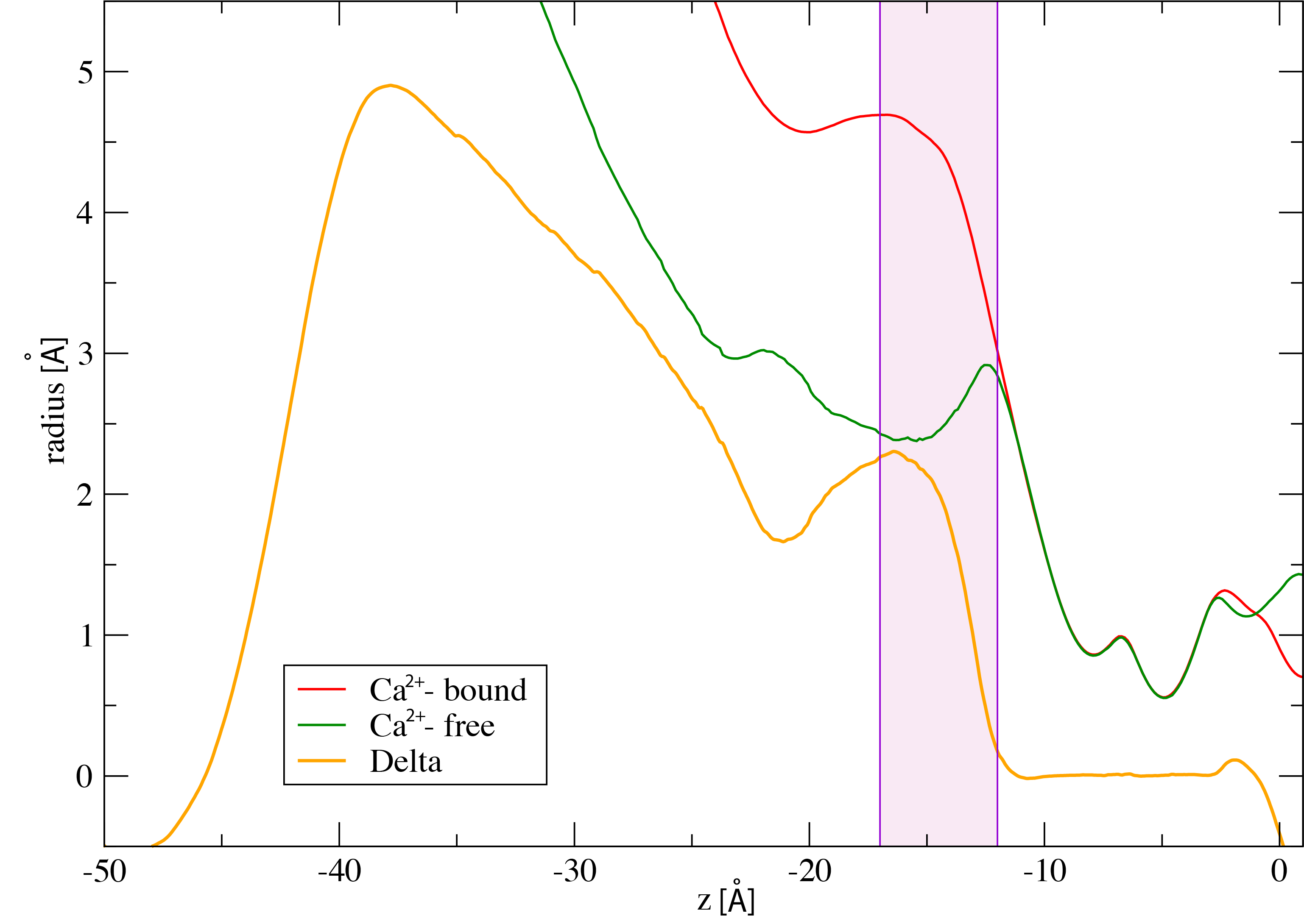}
\caption{\textbf{a)}. Side view of the transmembrane portion in the \Ca-free state. The blue sphere indicates the selectivity filter center of mass (z=0 \AA), the white spheres are located at z=-10,-20,-30 \AA, and the pink ones at z=-12,-17 \AA. For clarity, some subunits are not shown.\\ \textbf{b)} Average radius profiles for \Ca-free and \Ca-bound state. In orange is shown the difference $\Delta$ between \Ca-bound and \Ca-free radii. Notice that beneath the selectivity filter, $\Delta$ is equal to zero and starts increasing moving towards negative value of z, i.e. $z \sim  -12 \,$ \AA, until it reaches a first maximum at $z \sim  -17 \,$ \AA. These boundaries are chosen to define the Deep Pore Volume (DPV), here highlighted with a pink background. 
\label{fig:dpv}
}
\end{figure}

%%%%%%%%%%%%%%%%%%%%%%%%%%%%%%%%%%%%%%%%%%
\newpage
\subsection*{FIGURE S2. - Helices S6 orientation}

\begin{figure}[h]
\centering
\includegraphics[width=0.35\textwidth]{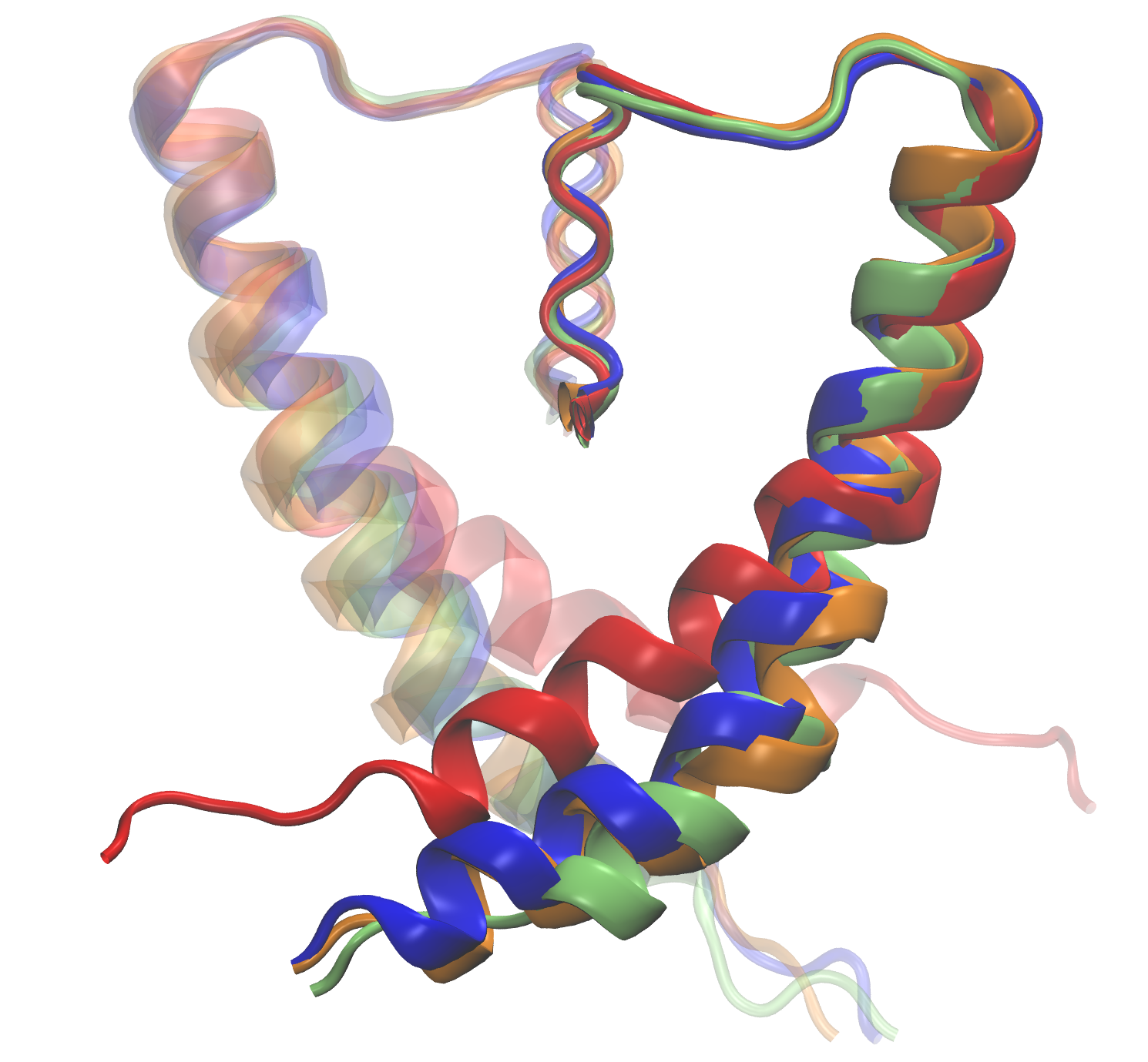}
\includegraphics[width=0.55\textwidth]{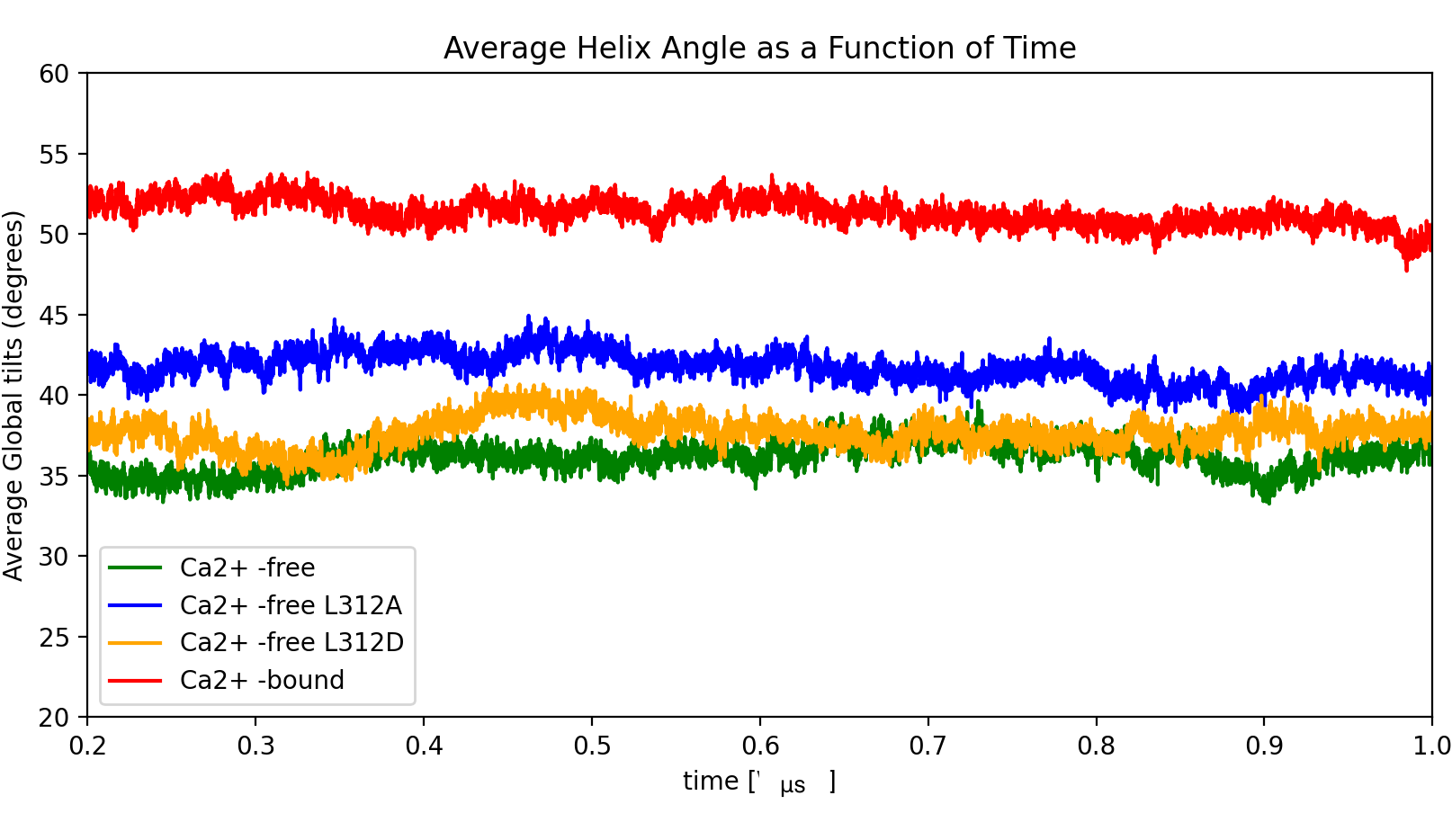}
\caption{
Right. Side view of the S6 helices for wild type \Ca-free (green), L312A \Ca-free (blue), L312D \Ca-free (orange) and wile type \Ca-bound (red). For clarity, only two subunits are shown. Left. Average angle between S6 and vertical axis as function of time. \\
We studied the S6 helices orientation with MDAnalysis HELANAL-routine~\cite{MichaudAgrawal2011,Gowers2016}, and averaged the S6 angle with the vertical axis.
The \Ca-free state results to be characterized by more elongated helices along the z-axis compared to \Ca-bound and the L312A mutant (Fig.~\ref{fig:helices}).
\label{fig:helices}
}
\end{figure}

%%%%%%%%%%%%%%%%%%%%%%%%%%%%%%%%%%%%%%%%%%
\newpage
\subsection*{FIGURE S3. - Helices S6 orientation statistics}
\begin{figure}[h]
\centering
\textbf{\Ca-free }\par\medskip \includegraphics[width=0.7\linewidth]{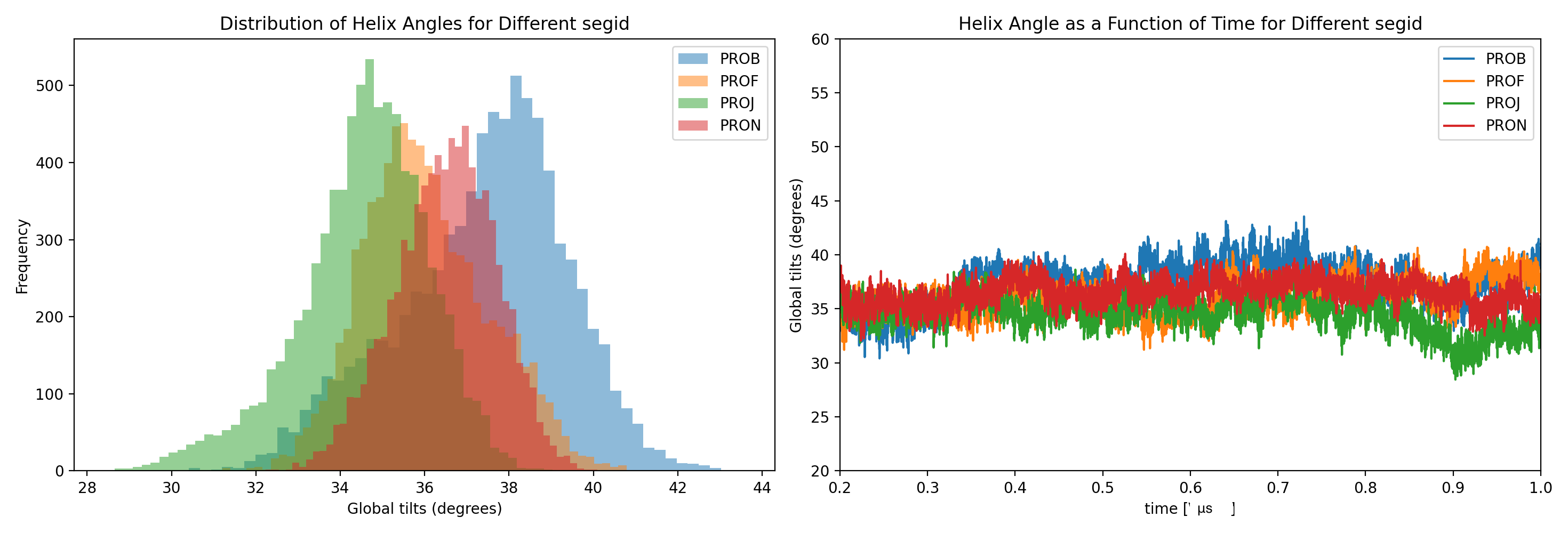} 
\end{figure}
\begin{figure}[h]
\centering
\textbf{\Ca-bound}\par\medskip \includegraphics[width=0.7\linewidth]{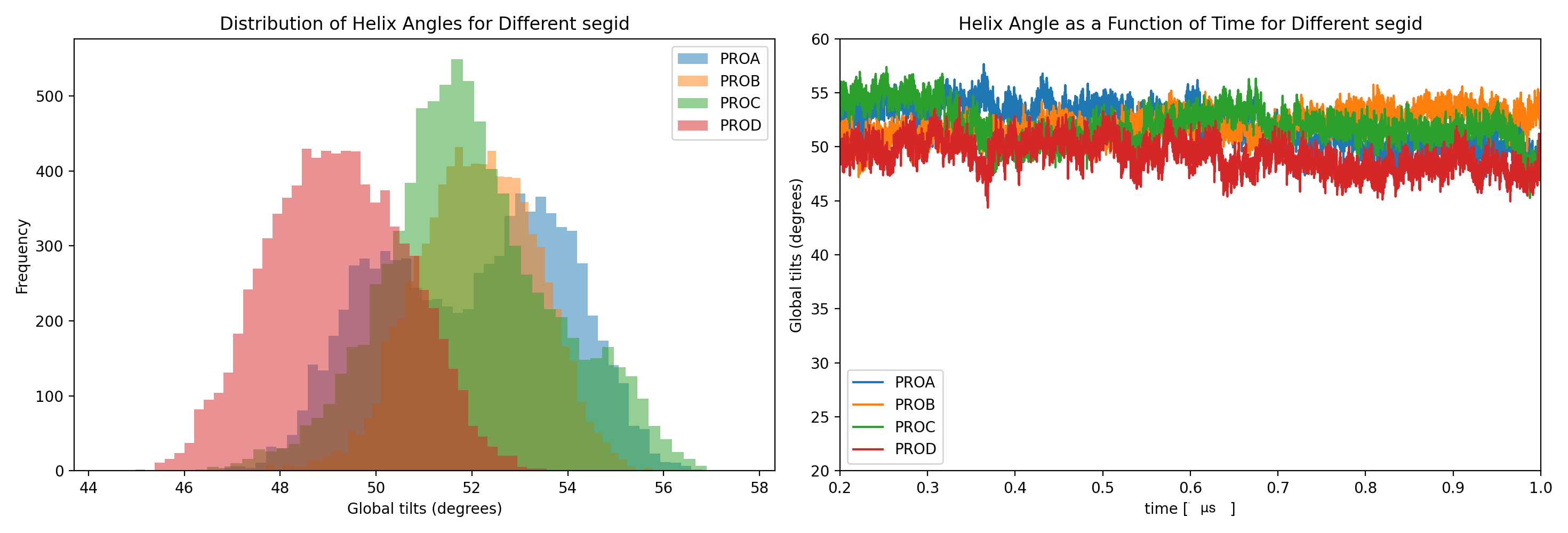}
\end{figure}
\begin{figure}[h]
\centering
\textbf{L312A}\par\medskip \includegraphics[width=0.7\linewidth]{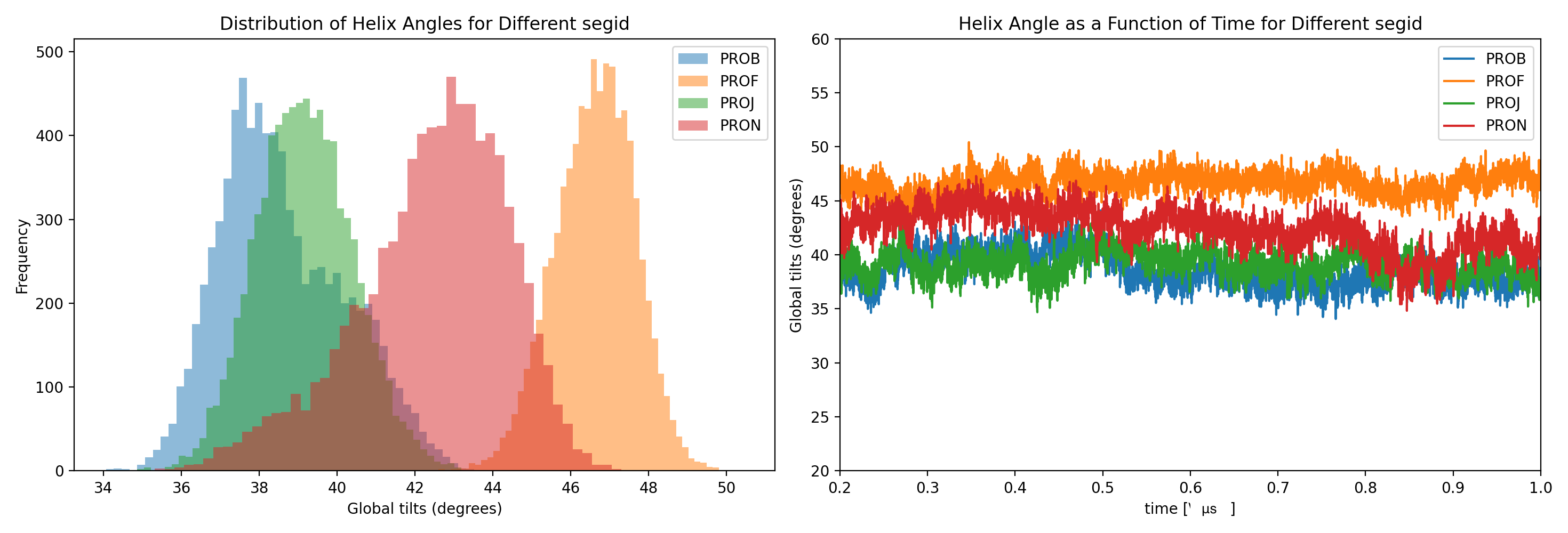}
\end{figure}
\begin{figure}[h]
\centering
\textbf{L312D}\par\medskip \includegraphics[width=0.7\linewidth]{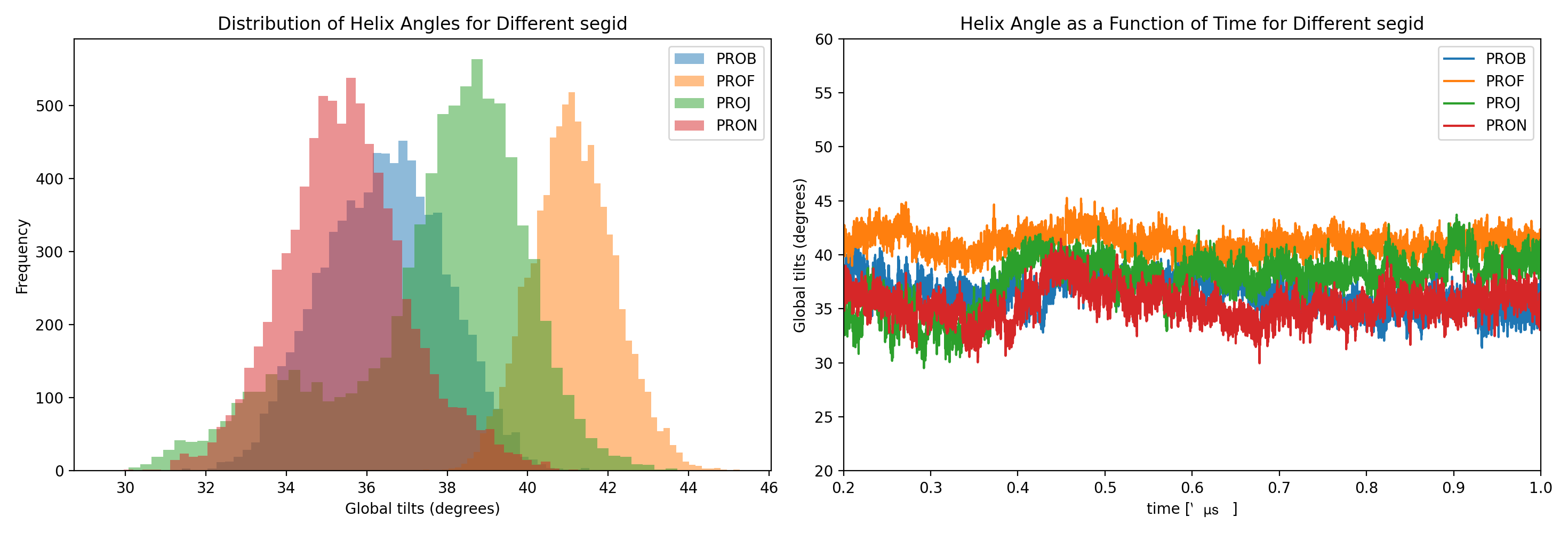}
\end{figure}
\begin{figure}[h]
\centering
 \begin{tabular}{lclclclc}
                \toprule
                \textbf{\Ca-free}  & & \textbf{\Ca-bound} & &  \textbf{L312A}  && \textbf{L312D}  \\
                \textbf{Mean} & \textbf{SD} & \textbf{Mean} & \textbf{SD} & \textbf{Mean} & \textbf{SD} & \textbf{Mean} & \textbf{SD} \\
                \midrule
                36.866 & 2.160  & 52.268 & 2.138 & 38.292 & 1.682 & 36.460 & 1.489    \\
                35.876 & 1.486  & 52.203 & 1.273 & 46.015 &1.847 & 40.948 & 1.197  \\
                34.639 & 1.567 & 52.063 & 1.750 & 38.973 & 1.449 & 37.023 & 2.596 \\
                36.947 & 1.781 & 49.337 & 1.593 & 42.228  & 1.968 &   35.081 & 1.728  \\
                \bottomrule
            \end{tabular}

    \caption{Summary statistics for S6-helix angles across different subunit and configurations: mean value and the sample standard deviation are listed for each case.
    \label{fig:helices_details}
    }    
\end{figure}

\newpage

\subsection*{FIGURE S4. - Lipid-protein interactions  }

\begin{figure}[h]
\centering
\includegraphics[width=\textwidth]{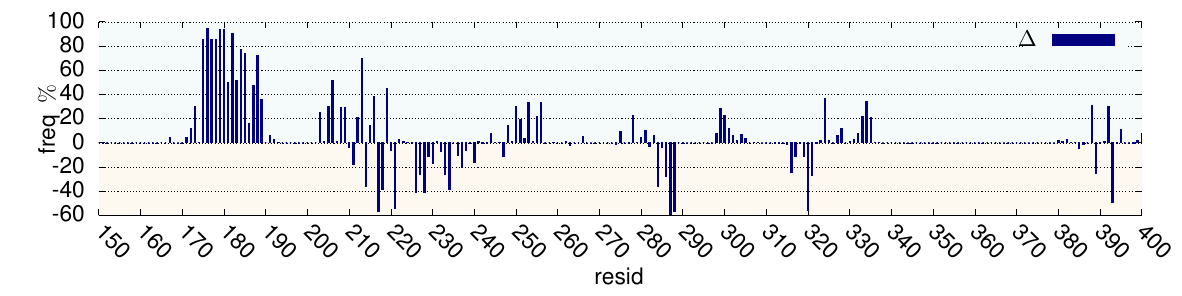}
\caption{The difference in interaction frequencies between the $^2+$-bound and Ca$^2+$-free states is expressed as a percentage of the time each residue spends in contact (within 5\AA) with lipids. For reference, S6 helices span from approximately residue 300 to 330, while S5 spans from 235 to 270. The blue background highlights where there is an excess of interactions for the \Ca-bound state, vice versa the yellow background shows interactions that are more frequent in the \Ca-free state compared to \Ca-bound state.
\label{fig:l-p-interaction}
}
\end{figure}

%%%%%%%%%%%%%%%%%%%%%%%%%%%%%%%%%%%%%%%%%%
\newpage
\subsection*{FIGURE S5. - K392 Displacement   }
\begin{figure}[h]
\centering
\includegraphics[width=\textwidth]{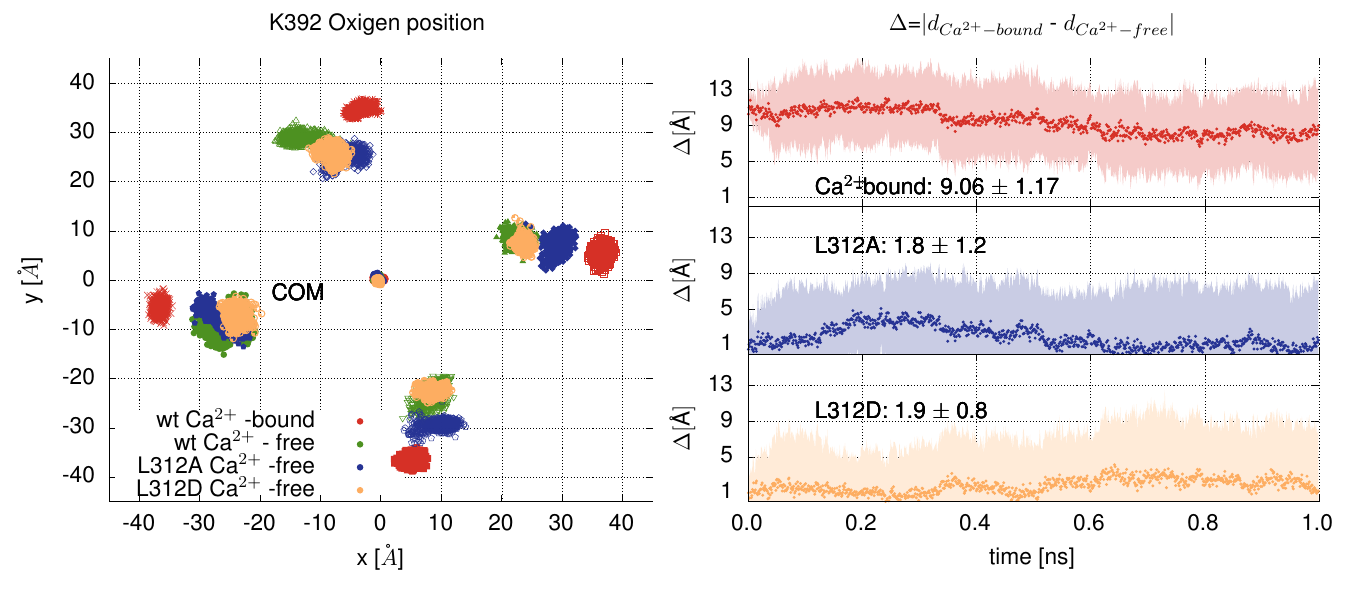}
\caption{ 
Left. K392-Oxygen atom xy coordinates for both \Ca-free, \Ca-bound state, and mutants. Right. Displacement of K392-Oxygen atom compared with the wild type \Ca-free.
\label{fig:displ}
}
\end{figure}

%%%%%%%%%%%%%%%%%%%%%%%%%%%%%%%%%%%%%%%%%%
\newpage
\subsection*{FIGURE S6. -  RMD trajectories}
\begin{figure}[h]
\centering
\includegraphics[width=1.0\textwidth]{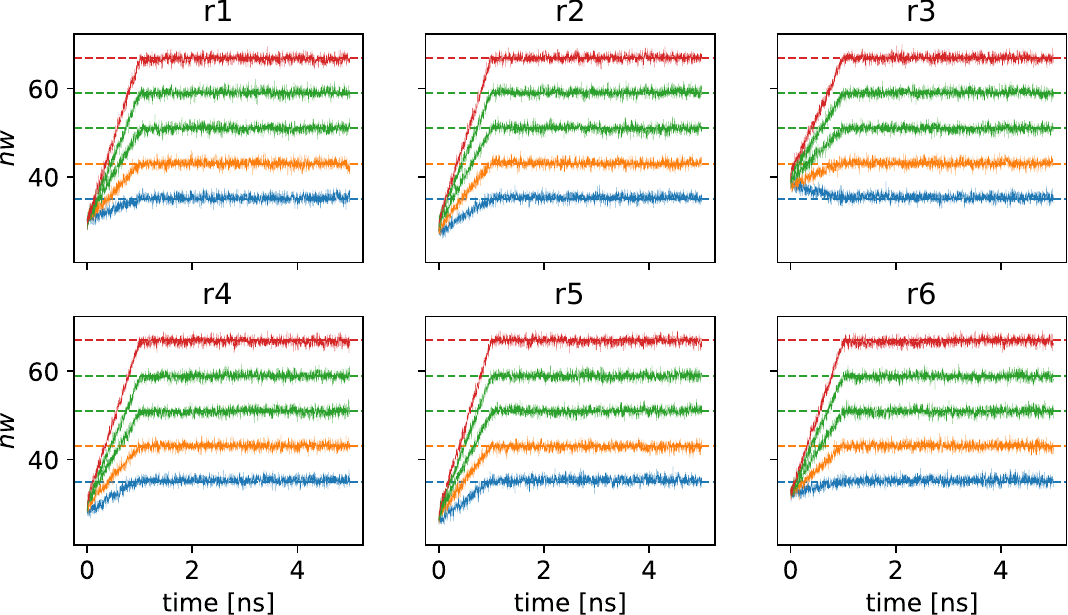}
\caption{
Number of water molecules $n_w$ sampled during some RMD simulations for the system shown in Fig. 4 of the main manuscript.
Each trace represent the sampling under a given constraint target $n_{w,0}$, represented by dashed lines.
The initial slope is the initial transient during which the target $n_{w,0}$ 
is updated from the initial unbiased value to the final one. 
Mean forces for each RMD are computed considering the sampling after 2 ns;
standard error of the mean force is estimated by using a block average procedure, 
with the length of each block being 100 ps.
Mean forces are integrated via simple euler method to obtain the free energy profile, 
and the error on the profile is computed by propagating the mean forces errors along the summation.
\label{fig:rmd_traj}
}
\end{figure}

%%%%%%%%%%%%%%%%%%%%%%%%%%%%%%%%%%%%%%%%%%
\newpage
\subsection*{FIGURE S7 -  RMD of the \Ca-free state without intruded lipids}
\begin{figure}[h]
\centering
\includegraphics[width=1.0\textwidth]{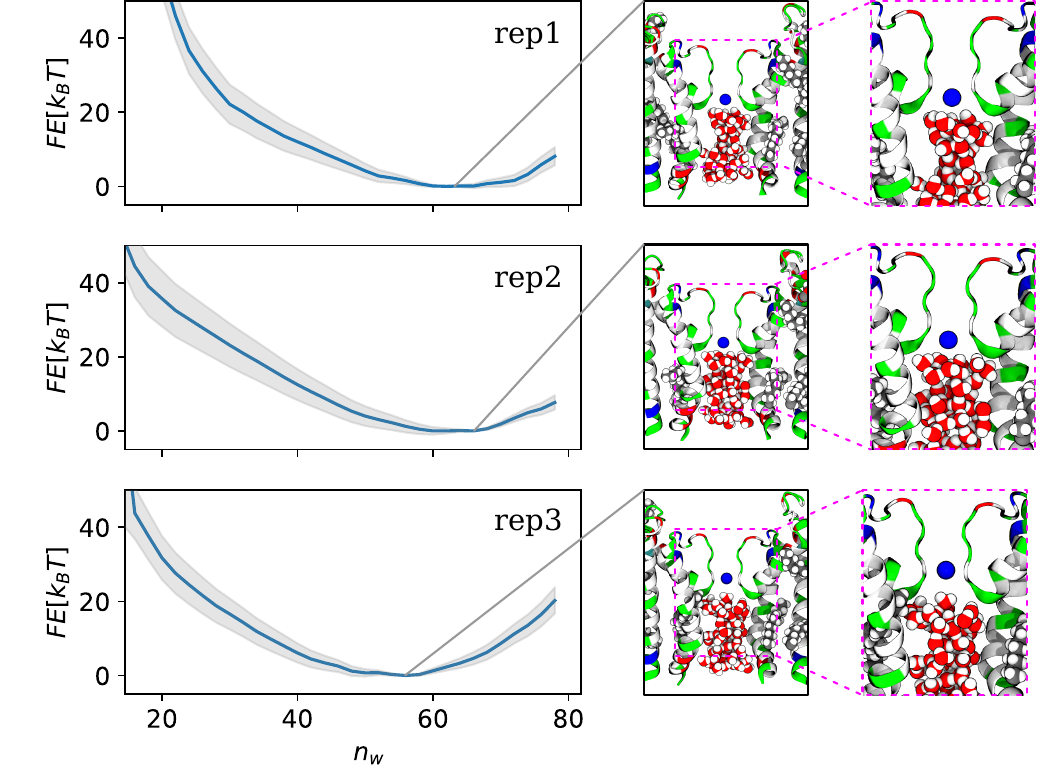}
\caption{Free energy profiles for three replica of the \Ca-free state before lipids intrude inside the DPV.
The profiles show that for all the three replica it exist only one global minimum, 
corresponding to the wet state, 
while dry states (lower filling level, $n_w < 45$) 
are found to be not stable.
This additional set of simulations largely support our hypothesis that lipid intrusion is the key feature to enable the pore gating.
Lateral panels represents the last frame of the RMD trajectory corresponding to minimum of the free energy profile. 
Protein is displayed in New Cartoon representation, while
water molecules and the potassium ion inside the selectivity filter are displayed in common VDW representation, rendered by using VMD~\cite{HUMP96}.
\label{fig:fe_nolipid}
}
\end{figure}

%%%%%%%%%%%%%%%%%%%%%%%%%%%%%%%%%%%%%%%%%%
\newpage
\subsection*{FIGURE S8 -  Compartison of TIP3P vs SPCE water model in RMD simulations}
\begin{figure}[h]
\centering
\includegraphics[width=1.0\textwidth]{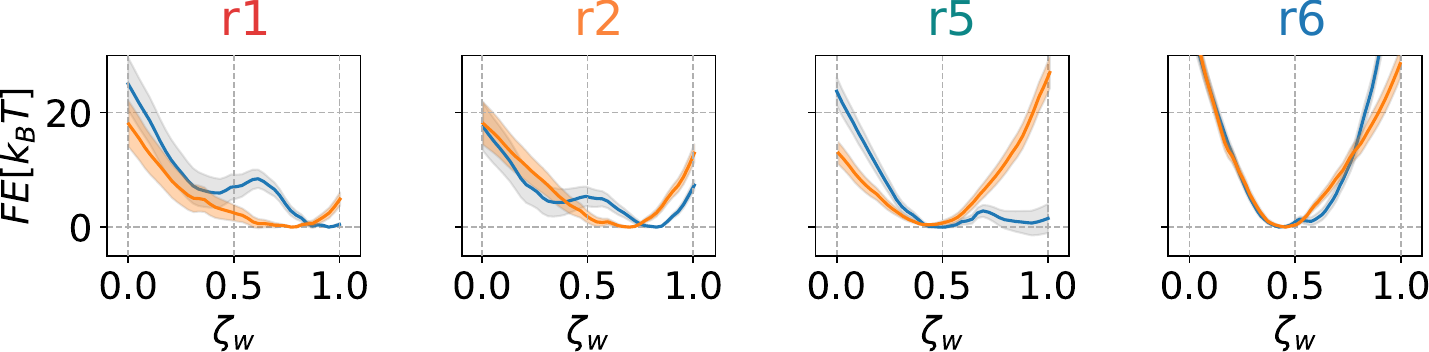}
\caption{Water filling free energy profiles computed via RMD simulations for the system reported in Fig. 4 of the main manuscript, computed also by using the TIP3P force field for the water, in combination with CHARMM36 force field for the protein and lipids, see orange lines. Blue lines are the same profile reported in the manuscript, representing the systems simulated with SPC/E water model in combination with Amber force field, see methods. 
The most notable difference lie in the fact that orange lines do not show metastabilities in any of the simulated replicas. This fact is not particularly surprising, since TIP3P do not well reproduce the surface tension of the water. Another difference is that in the wet state TIP3P seems to display a lower density of water inside the control box. This is in line with previous reported works, where the lowering of density of water was correlated to the lower conductance of the closed state~\cite{gu2023central}.
Neverthless, TIP3P predicted a wet minimum for replica $r1$ and $r2$, while a dry one for replicas $r5$ and $r6$,
overlall confirming the observation that the hydrophobic gating can easily happen in presence of lipid tail intruded into the DPV. 
\label{fig:tip3pvsspce}
}
\end{figure}

%%%%%%%%%%%%%%%%%%%%%%%%%%%%%%%%%%%%%%%%%%

\end{document}